\documentclass[aps,prl,preprint,floatfix,showpacs,groupedaddress,superscriptaddress,longbibliography]{revtex4-1}

\usepackage{multirow}
\usepackage{graphicx}

\begin{document}

\title{InAs nanowire with epitaxial aluminium as a single-electron transistor with fixed tunnel barriers}

\author{M.~Taupin}
 \thanks{Present address: Institute of Solid State Physics, TU Wien, Wiedner Hauptstr. 8-10, 1040 Vienna, Austria}
 \email{mathieu.taupin@ifp.tuwien.ac.at}
 \affiliation{Low Temperature Laboratory, Department of Applied Physics, Aalto University School of Science, P.O. Box 13500, FI-00076 Aalto, Finland}
 \author{E.~Mannila}
 \affiliation{Low Temperature Laboratory, Department of Applied Physics, Aalto University School of Science, P.O. Box 13500, FI-00076 Aalto, Finland}
\author{P.~Krogstrup}
  \affiliation{Center For Quantum Devices, Niels Bohr Institute, University of Copenhagen, Universitetsparken 5, 2100 Copenhagen \O, Denmark} 
\author{V.~F.~Maisi}
  \affiliation{Low Temperature Laboratory, Department of Applied Physics, Aalto University School of Science, P.O. Box 13500, FI-00076 Aalto, Finland} 
  \affiliation{Center For Quantum Devices, Niels Bohr Institute, University of Copenhagen, Universitetsparken 5, 2100 Copenhagen \O, Denmark} 
\author{H.~Nguyen}
 \affiliation{Low Temperature Laboratory, Department of Applied Physics, Aalto University School of Science, P.O. Box 13500, FI-00076 Aalto, Finland}
 \affiliation{Center For Quantum Devices, Niels Bohr Institute, University of Copenhagen, Universitetsparken 5, 2100 Copenhagen \O, Denmark}
 \affiliation{Nano and Energy Center, Hanoi University of Science, VNU, 120401 Hanoi, Vietnam}
\author{S.~M.~Albrecht}
  \affiliation{Center For Quantum Devices, Niels Bohr Institute, University of Copenhagen, Universitetsparken 5, 2100 Copenhagen \O, Denmark}
  \author{J.~Nyg\r{a}rd}
  \affiliation{Center For Quantum Devices, Niels Bohr Institute, University of Copenhagen, Universitetsparken 5, 2100 Copenhagen \O, Denmark}  
\author{C.~M.~Marcus}
  \affiliation{Center For Quantum Devices, Niels Bohr Institute, University of Copenhagen, Universitetsparken 5, 2100 Copenhagen \O, Denmark}
\author{J.~P.~Pekola}
 \affiliation{Low Temperature Laboratory, Department of Applied Physics, Aalto University School of Science, P.O. Box 13500, FI-00076 Aalto, Finland}

\date{\today}

\begin{abstract}
We report on fabrication of single-electron transistors using InAs nanowires with epitaxial aluminium with fixed tunnel barriers made of aluminium oxide. The devices exhibit a hard superconducting gap induced by the proximized aluminium cover shell and they behave as metallic single-electron transistors. In contrast to the typical few channel contacts in semiconducting devices, our approach forms opaque multichannel contacts to a semiconducting wire and thus provides a complementary way to study them. In addition, we confirm that unwanted extra quantum dots can appear at the surface of the nanowire. Their presence is prevented in our devices, and also by inserting a protective layer of GaAs between the InAs and Al, the latter being suitable for standard measurement methods.
\end{abstract}

\maketitle
\section{Introduction}
Semiconducting nanowires (NWs) are widely used nowadays in nanotechnology \cite{Samuelson2004,Thelander2006,Lieber2007} as their transport properties can be easily tuned \cite{Doh2005,Liang2015}. In particular, InAs NWs are of interest as they are optically active \cite{Li2006} and can act as a field-effect transistor \cite{Bryllert2006}, a quantum dot \cite{Bjork2004,Dam2006,Sand-Jespersen2007,Lee2012,Chang2013,Lee2014} or a qubit \citep{Nadj-Perge2010}. Recently, the growth of a NW with a high quality interface between InAs and aluminium has been achieved \cite{Krogstrup2015}, with a hard superconducting gap \cite{Chang2015}, systems in which Majorana bound states have been observed \cite{Lutchyn2010,Oreg2010,Das2012,Albrecht2016}. In these devices, the barriers are formed electrostatically to allow great flexibility of the barrier strength, contrary to ``fixed'' tunnel barriers one can find in metallic single-electron transistors (SETs). However the drawback of this flexibility is the limited number of open conductive channels \cite{Albrecht2016,Higginbotham2015} which can either limit the signal in case of large opacity of the barriers, or induce a leakage current in the other limit.

In this study, we present a simple device in an InAs NW proximized with epitaxial Al. The main idea is to use the aluminium shell on top of the InAs NW to form a fixed tunnel barrier, thus with InAs as a SET, as in a metallic system \cite{Fulton1987,Grabert1992}. Compared to electrostatic tunnel barriers, aluminium oxide based tunnel contacts are known to possess superior properties: They have a large number of conduction channels, typically of the order of $10^4$~\cite{Averin2008,Maisi2011}. It allows one to make them several orders of magnitude more opaque than the few channel contacts without losing in signal strength. The more opaque the tunnel junctions are, the better the approximation of sequential tunnelling is. Hence when probing the hardness of the superconducting gap, we observe consistently an order of magnitude lower leakage levels in the gap. For this reason, the combination of a metallic SET and a proximized InAs NW can give access to functionalities mixing SET and InAs NWs properties, not possible with standard techniques. In addition, our method reduces some technical difficulties: Only one gate per intentional QD is needed and it ensures good contacts between the NW and the external leads. It also prevents the appearance of parasitic effects due to the exposure of the InAs core during the fabrication process as the InAs core as well as the interface between InAs and Al remain intact. Such effects are prevented as well by inserting a protective layer between the InAs core and the Al layer. Several works already exist on the fabrication of a SET using semiconducting NWs with fixed tunnel barriers (i.e. not tunable by gate modulation), with, e.g., Si NWs \cite{Zhong2005,Hofheinz2006} or InAs/InP heterostructures \cite{Thelander2003}.
\\

\section{Fabrication}

\begin{figure}
	\centering
		\includegraphics[width=0.75\textwidth]{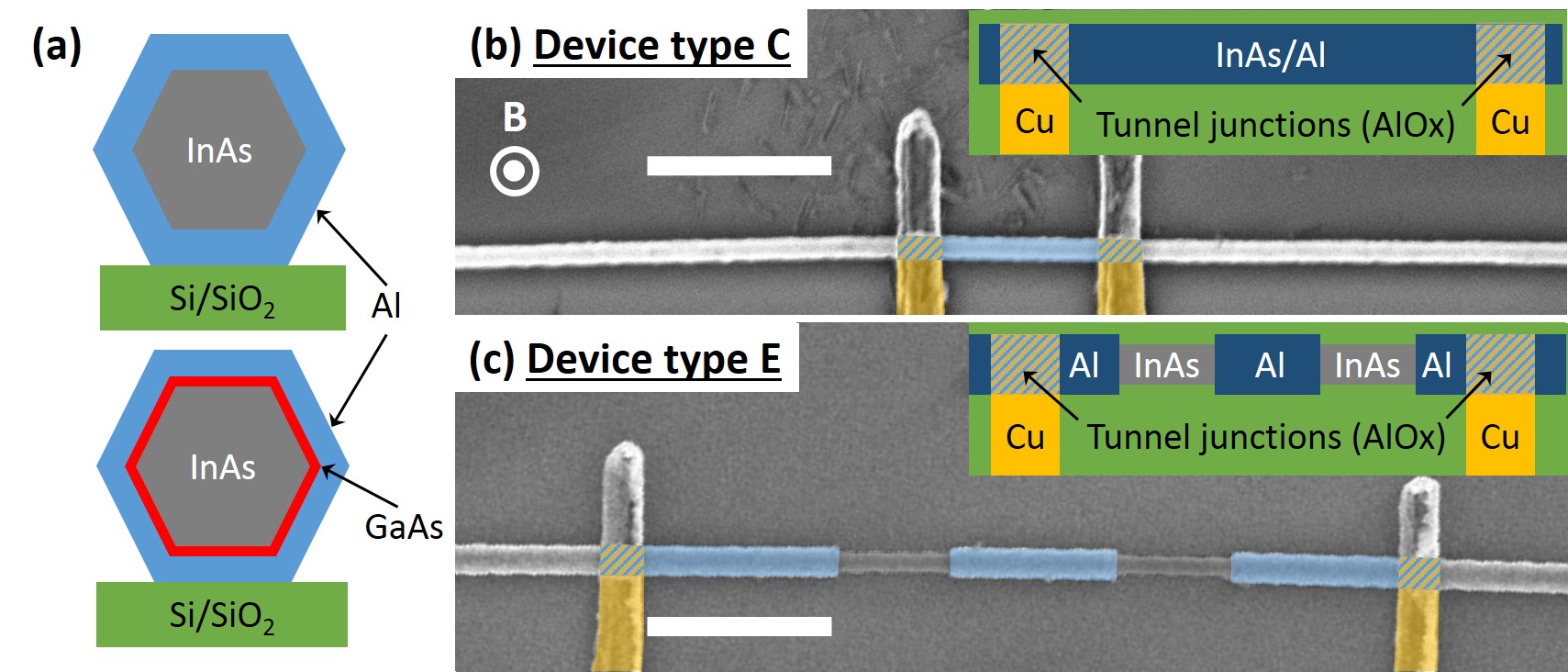}
	\caption{Panel \textbf{(a)}: Sketch of the cross-section of the NW. The aluminium shell (symbolized in blue) is grown epitaxially either directly on the InAs wire, in gray (upper sketch, devices named ``\hbox{-Al}''), or on an intermediate protective GaAs shell (lower one, named ``\hbox{-GaAs}''). Panels \textbf{(b)} and \textbf{(c)}: Sketch and SEM image of a device type C, with a fully covered wire, and of a device type E, with an isolated aluminium island. The leads and the side gate (not shown) are made of copper (in orange), with a thickness from 100 to 200~nm and the tunnel barriers are formed of aluminium oxide and are situated on the hatched surfaces. The not-drawn areas on the devices are supposed to play no role in the properties of the system. The white bars represent 1~$\mu$m.}
	\label{fig1}
\end{figure}

The hexagonal InAs NWs are grown by molecular beam epitaxy (MBE) using gold nanoparticles catalyst and are $10-15$~$\mu$m long. The aluminium is then deposited epitaxially, covering entirely the NW, without breaking the vacuum to guarantee a good interface between InAs and Al \cite{Krogstrup2015}. For some NWs, a buffer layer of GaAs, $5$~nm thick, is grown on top of the InAs, followed then by the Al deposition. These NWs form a stacking-fault free wurtzite phase, with misfit dislocations at the InAs/GaAs interface due to the 7\% lattice mismatch, therefore the strain relaxes very quickly \cite{Popovitz-Biro2011}. This intermediate GaAs layer is expected to reduce the stress at the surface of the InAs and improve the intrinsic properties of the NW (like e.g. carrier mobility), as already observed in various NWs with cover shells \cite{Tilburg2010,Chia2012,Lin2013,Holloway2013,Ganjipour2014}. A sketch of the cross section of the wires is shown Fig. \ref{fig1}\textbf{(a)}. The devices with the GaAs covered shell are named ``\hbox{-GaAs}'', and the others, with only the aluminium shell, are named ``\hbox{-Al}''. We remind that the NWs with the GaAs layer also have an epitaxial layer of Al. We have a clean contact between the core and the Al layer which leads to a negligible energy barrier (i.e. no tunnel barrier is formed between the NW core and the superconducting layer) as shown previously \cite{Krogstrup2015, Chang2015}. 

The NWs are transferred from the growth chip on a pre-marked substrate by dry deposition. The substrate is a highly-doped silicon wafer covered by 200~nm of silicon oxide and is used as a backgate. The position of the NWs is found on the chip using a scanning electron microscope (SEM). In order to study the effect of the chemical etching, we first isolate an Al island ($\sim 1~\mu$m long) in the middle of the NW, by etching chemically at two places a 0.5 to 1~$\mu$m segment of the Al shell by immersion in MF-CD-26 for 90~s at room temperature (called device ``E'', for Etched). The remaining Al on the central island and on each side of the wire, close to the junctions, is supposed to keep the proximized superconductivity intact and uniform over the entire NW. The other type of devices is made without the chemical etching, keeping the aluminium shell intact and thus without any bare InAs (called device ``C'', for Covered). The pre-marked chip is then covered with a resist and the two leads and the side gate are patterned by electron-beam lithography. After development, the pre-marked chip is inserted in an electron-beam evaporator equipped with a plasma gun. The native oxide layer on Al is removed by argon plasma etching inside the evaporator chamber. The epitaxial Al is then re-oxidised under O$_2$ atmosphere of 2~mbar for 2~minutes to create the tunnel barriers, approximately 0.5 to 1~nm thick \cite{Gloos2003}. 150 to 200~nm of Cu is next evaporated in order to make the leads and the side gate. The fact that the native Al oxide is etched \textit{in-situ} ensures good control of the tunnel junctions. The tunnel barriers are expected only at the junctions and no barrier should form inside the NW. The junctions cover from 150 to 300~nm over the wire depending on the device and are spaced by 1~$\mu$m (for the device C-Al') and 5~$\mu$m for the others. During the plasma etching and the oxygen re-oxidation, the InAs core is protected either by the Al shell where the junctions are made and by the resist everywhere else. Therefore we do not expect these treatments to damage further the InAs core. We emphasize that only Cu is deposited on the pre-marked chip, i.e. no Al layer is added: The tunnel junctions are formed by re-oxidizing the epitaxial Al layer grown from the MBE process after etching the native oxide layer. The sketches and SEM images of the devices are shown Figs. \ref{fig1}\textbf{(b)} and \textbf{(c)}. The main parameters of the samples are given in Table \ref{tab1}. The NWs from the devices E-Al and C-Al come from the same growth, and the same applies for the devices E-GaAs and C-GaAs. The smaller resistance and charging energy of the device E-GaAs come from its wider tunnel junctions compared to the other devices. Only the backgate was used in this study, but we have obtained similar results using the side gate. All the measurements presented here have been performed in a dilution fridge at a bath temperature $T_{bath}\simeq60$~mK and, when applied, the magnetic field is perpendicular to the NW (see Fig. \ref{fig1}\textbf{(b)}). In the last section, conductance measurements have been performed, a lock-in amplifier was used with the excitation voltage ranging from $V_{ac}=2$ to 10~$\mu$V and the frequency from $f\approx0.7$ to 300~Hz depending on the gain and the bandwidth of the voltage and current amplifiers used. 
 
According to Ref. \cite{Wunnicke2006}, the capacitance of the NW on the highly doped Si substrate is estimated to be $C_{g}\sim0.25$~fF for each device. We can then estimate the capacitance of the junctions supposing they cover half of a cylindrical NW (we cannot evaporate below the NW). In the case of aluminium oxide, we use the dielectric constant $\epsilon_r\approx4$ and the oxide thickness between 0.5 and 1~nm \cite{Gloos2003}. We obtain thus the capacitance per junction $C_J\sim1$~fF, and thus the total capacitance $C_{\Sigma}=C_g+2C_J\sim2.25$~fF, which gives the charging energy $E_C=e^2/(2C_{\Sigma})\sim35$~$\mu$eV, close to the experiment. The total capacitance of the devices is mainly caused by the junctions, which are rather wide in our devices.
\\

\begin{table}
	\centering
	\begin{tabular*}{0.75\textwidth}{@{\extracolsep{\fill}} c c c c c c c}
		\hline \hline
		\multirow{2}{*}{Name} & \multirow{2}{*}{$R_{T}$} & $\Delta$ & $E_c$ & $\phi_{\text{InAs}}$ & $t_{\text{Al}}$ & $t_{\text{GaAs}}$ \\	
		 & & ($\mu$eV) & ($\mu$eV) & (nm) & (nm) & (nm)\\
		\hline
		E-Al & 180~k$\Omega^a$ & $130^a$ & 40$^a$ & \multirow{2}{*}{70} & \multirow{2}{*}{20} & \multirow{2}{*}{n/a}\\
		C-Al & 224~k$\Omega$ & 202 & 45 & &\\
		C-Al' & 126~k$\Omega$ & 192 & 30 & 40 & 20 & n/a \\
		\hline
		E-GaAs & 64~k$\Omega$ & 195 & 10 & \multirow{2}{*}{50} & \multirow{2}{*}{25} & \multirow{2}{*}{5}\\
		C-GaAs & 19~M$\Omega$ & 195 & 65 & &\\
		\hline \hline
	\end{tabular*}
	\raggedright $^a$ Taken at $V_{BG}=2$~V
	\caption{Summary of the main parameters of devices, obtained by fitting the normal state at $T_{bath}$: the total resistance across the whole device at low temperature $R_{T}$, the superconducting gap $\Delta$, the charging energy $E_c$, the diameter of the InAs core $\phi_{\text{InAs}}$, the thickness of the aluminium layer $t_{\text{Al}}$ and of the GaAs layer $t_{\text{GaAs}}$. The device C-Al' is similar to the device C-Al with a different size.}
	\label{tab1}
\end{table} 

\section{Characterization of the devices}
\begin{figure}
	\centering
		\includegraphics[width=0.75\textwidth]{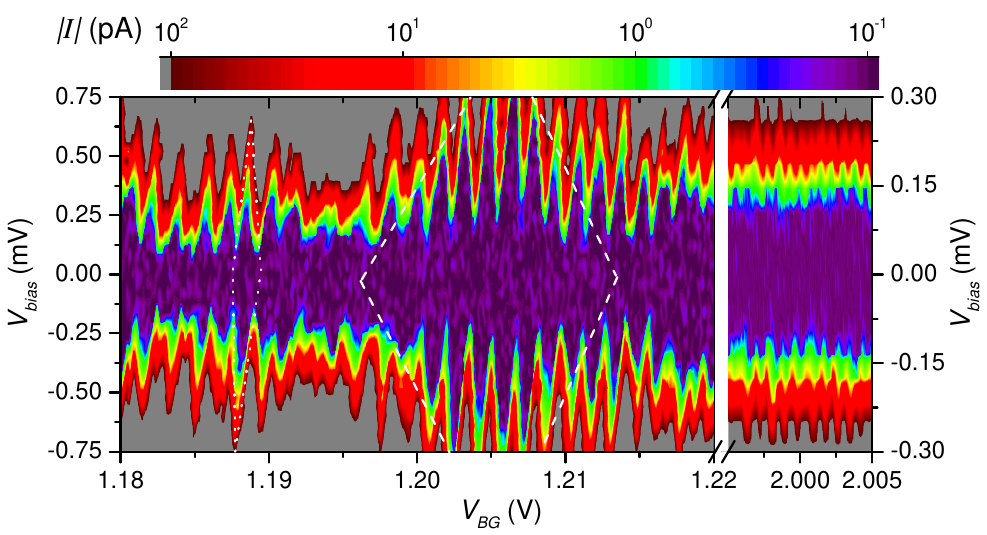}
	\caption{Current $I$ vs bias $V_{bias}$ and backgate $V_{BG}$ centred at 1.2 and 2~V, respectively. Several sets of Coulomb diamonds are visible, with different gate periodicity and amplitude, see e.g., the white dashed and white dotted diamonds. Note the different vertical scales between the left and the right part of the graphs.
	}
	\label{fig2}
\end{figure}

We first study the device E-Al, Fig. \ref{fig1}\textbf{(c)}, in which the InAs core is exposed. Here, we choose the wire without GaAs, and etch Al in selected areas. The effect of the backgate on electron transport measurements is shown in Fig. \ref{fig2}. At negative backgate values $V_{BG}<0$ (not shown), the transport is blocked. For $0<V_{BG}<2$~V, a complex stability diagram is present, with at least two sets of Coulomb diamonds (see e.g. the white dotted and dashed diamonds in the left-hand side of Fig. \ref{fig2}). Only one QD is expected with a small charging energy. However, the QDs measured at intermediate gate values, around 1.2~V, show a charging energy between 0.3 and 0.8~meV, too high to reflect the main dot. These large values as well as the aperiodicity of the diamonds with $V_{BG}$ are signatures of the presence of several QDs. Similar behaviour has been observed in several of our devices of the same type and unwanted QDs have as well been reported in previous study, despite the high quality of the NWs used \cite{Bleszynski2007,Boyd2011,Chang2015}. It is believed to be caused by defects \cite{Schroer2010} and potential fluctuations at the surface of the NW \cite{Weis2014} which can be triggered by chemical and/or plasma treatments of the NW. One reason for their presence comes from the fabrication process: The contacts on the NW are made directly on it, and in order to have an ohmic contact or to etch away a surface layer (the Al epitaxial layer in our case, see, e.g. the devices in Refs.~ \cite{Albrecht2016,Higginbotham2015}), additional chemical or plasma cleaning of the surface may be needed. These processes may deteriorate the surface of InAs, increasing thus the likelihood of forming unwanted QDs whose locations and sizes are not controlled. It is nevertheless possible to make them transparent by tuning locally its potential with additional side gates, leading to the fabrication of complex devices with potentially unnecessary side gates (see, e.g., some of the devices in Ref. \cite{Albrecht2016}). This is achieved in our situation by increasing the backgate voltage: Above $V_{BG}\sim2$~V, although deformed, the stability diagram is more regular and periodic with $V_{BG}$ (right-hand side of Fig. \ref{fig2}), similar to a metallic device. This corresponds to the intended dot with a charging energy of $\sim40~\mu$eV. The superconducting gap is, however, small in this device compared to the other ones (see Table \ref{tab1}). One possible reason for this is that the extra QDs affect the superconducting state of the NW. 
\\

\begin{figure}
	\centering
		\includegraphics[width=0.75\textwidth]{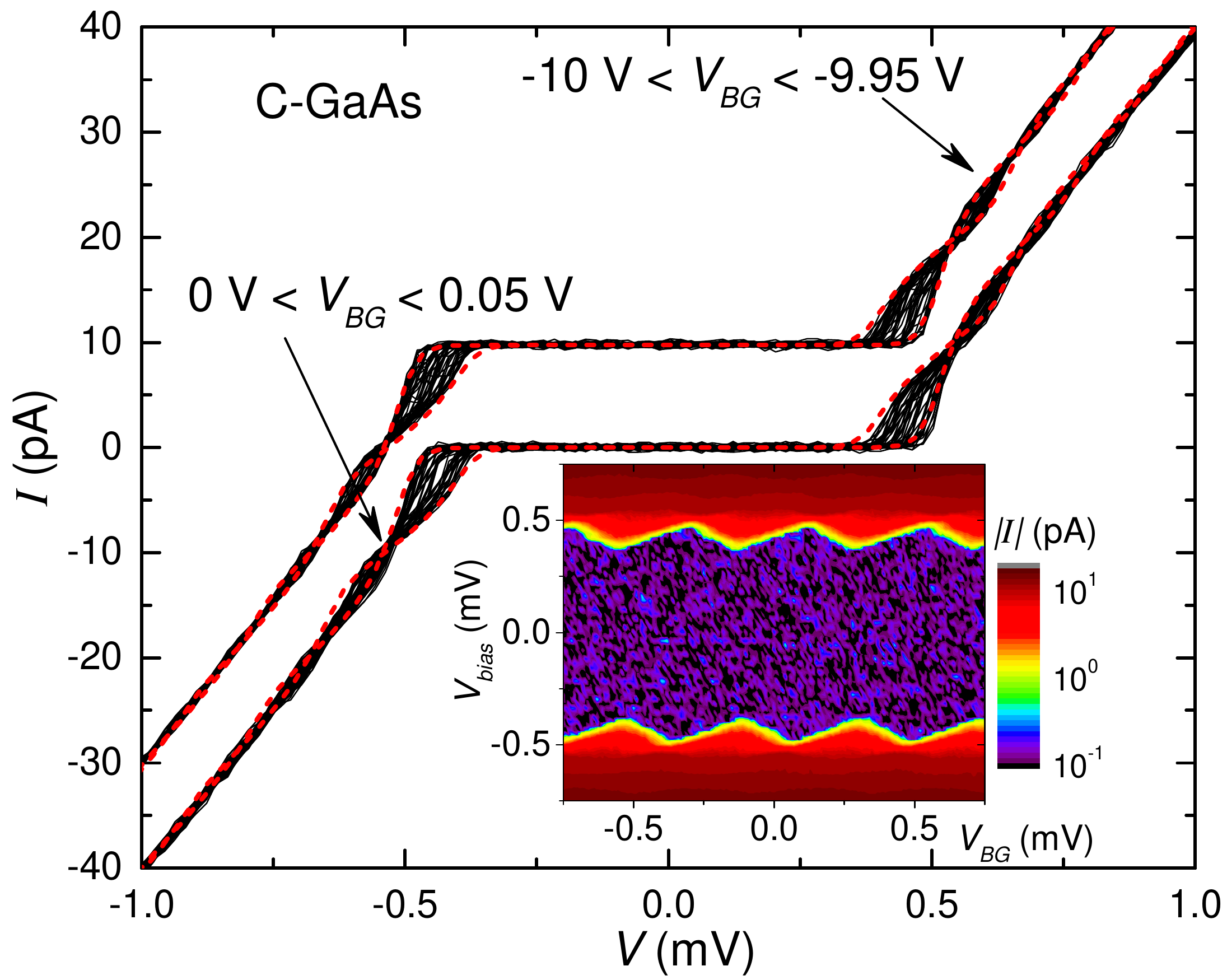}
	\caption{I-V characteristics of the device C-GaAs taken at several backgate positions, close to $-10$~V (upper curve) and 0~V (lower curve). The upper curve has been shifted by $10$~pA for visibility. The dashed lines correspond to theoretical fits, with the same parameters for both measurements. The inset is the stability diagram centred at $V_{BG}=0$~V.}
	\label{fig3}
\end{figure}

In our other devices, the InAs core is unexposed and always covered by another layer, either by the Al shell (device C-Al), by a protective GaAs shell (device E-GaAs) or by both (device C-GaAs). In the linear ohmic regime, at large bias voltage values, these devices exhibit a metallic behaviour: The transport is independent of the backgate value (no noticeable differences have been seen for $V_{BG}$ in the range $-5$ to 5~V). The I-V characteristics of the device C-GaAs at the backgate positions close to $-10$~V and 0~V are shown in Fig. \ref{fig3}. Both sets of I-V characteristics are similar, confirming the metallic-like state of our device. The dashed lines correspond to a theoretical fit of the normal state used for a metallic SET with a SC island \cite{Grabert1992}: The agreement between the measurements and the fit is very good. The parameters used for the fit are given in Table \ref{tab1} and are the same for both measurements. The theoretical model fits also nicely the I-V characteristics of the other devices (not shown) in the normal state. This indicates that the QD measured is different from the one measured in the device E-Al as the properties of the devices (total resistance, charging energy and superconducting gap) do not change with the gate voltage. The inset is the stability diagram around $V_{BG}=0$~V, exhibiting periodic and regular Coulomb diamonds. The metallic behaviour (i.e. absence of gate dependence in the transport besides the Coulomb blockade regime) is an evidence of the absence of undesirable QDs and that the transport is governed by only one intrinsic QD.

\begin{figure}
	\centering
		\includegraphics[width=0.75\textwidth]{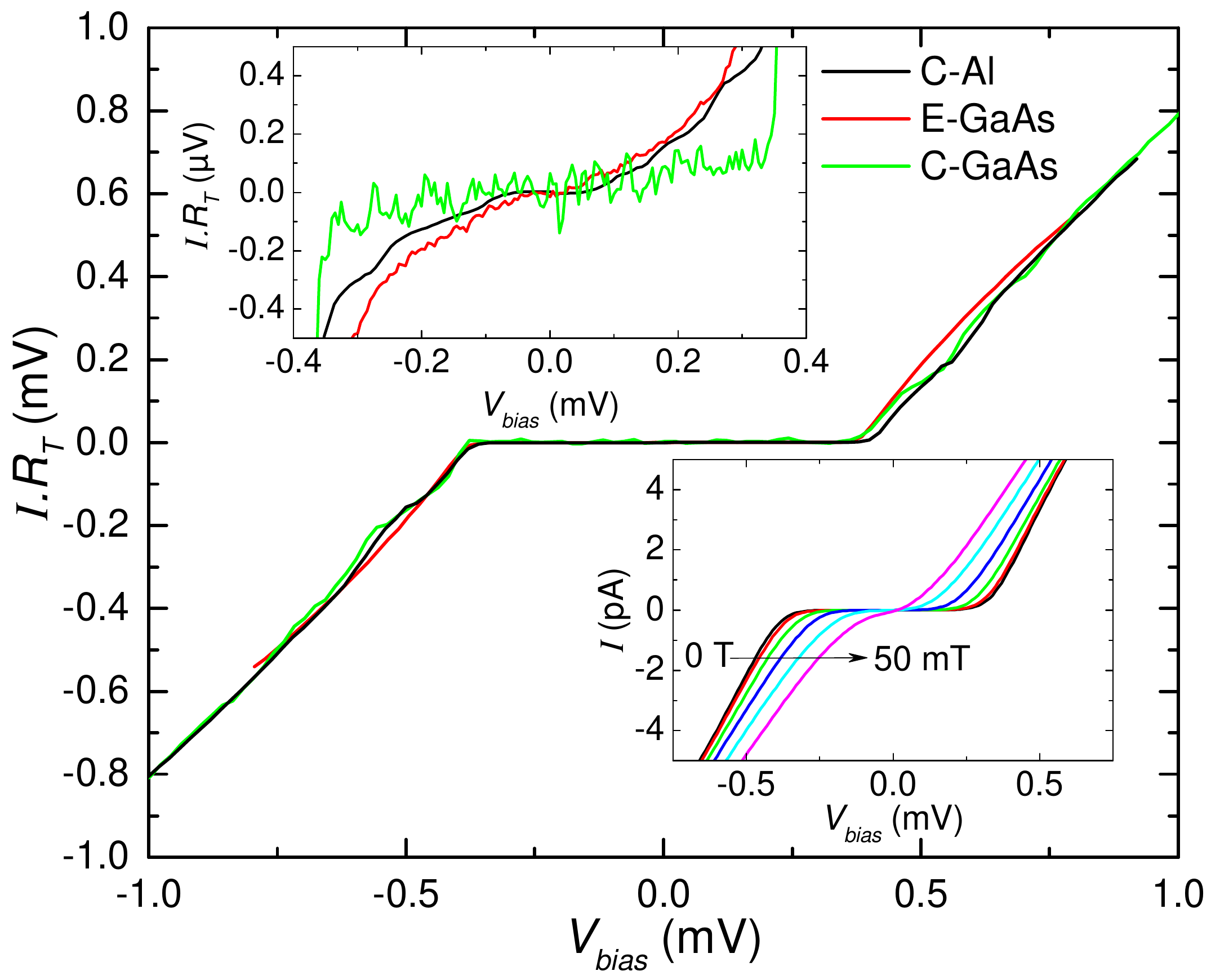}
	\caption{I-V characteristics of the device C-Al, E-GaAs and C-GaAs. The vertical axis is the product of the current and the resistance of the device $I \cdot R_T$ to normalize the measurements. The upper inset is a magnification in the subgap state, and the lower one the field dependence of a device similar to C-Al, from $B=0$~mT to 50~mT with 10~mT step. The measurements are done in the gate open state.
	}
	\label{fig4}
\end{figure}

The comparison of the transport measurements at low bias voltage in the gate open state of the three devices is shown Fig. \ref{fig4}. To present all the samples on the same footing, we plot the product of the current and the resistance $I \cdot R_T$. The three devices are similar with the main difference being the charging energy. The upper inset shows the magnification of the measurements in the superconducting state. All devices have a superconducting gap similar to that of Al, $\sim 200~\mu$eV. The slope of the I-V characteristics in the superconducting state is a signature of the hardness of the gap, and the ratio between the conductance in the superconducting state $G_S$ and in the normal state $G_N$ is $G_S/G_N\lesssim 10^{-3}$, a measure of the hard gap of our system. As we have opaque transport channels, we now obtain an order of magnitude lower ratios proving that the gap in the NWs is even an order of magnitude harder than estimated earlier in Ref.~\cite{Chang2015}. The hard gap is not affected by etching the Al shell in the device E-GaAs, since the value measured equals to the gap at the proximity of the junctions, where the Al shell is not etched chemically. The lower inset shows the field dependence of a device similar to C-Al, from $B=0$~mT to 50~mT, close to the critical field measured at $B_{c2}\approx55$~mT. The main effect of the magnetic field in the region $|eV_{bias}|\geq2\Delta(B)$ is to close the superconducting gap: From this point of view, our devices are identical to metallic SETs and do not seem to present any additional interest. We will thus not focus on the normal state under field any longer.
\\

The devices we show can be used for future studies of the properties of proximized superconductivity as our method is relatively non-invasive. Until now the SET regime in proximized InAs NWs was only achieved by tuning the potential of the wire with gates \cite{Albrecht2016,Higginbotham2015}. However, the transport properties of the system can be very sensitive to the gate positions, and corrections have to be applied in case of cross-talk between the leads and the gates or between each gate. With one gate only, this is not the case, and we thus have a possibility to perform more advanced experiments, such as using the devices as a turnstile \cite{Pekola2008}. The charging energy of our device can be easily increased by decreasing the dimensions of the NW (total length and diameter), the size of the QD (junctions spacing) or the size of the junctions.

The similarity between the devices C-Al and C-GaAs suggests that NWs with GaAs cover shell can be used for an SET setup, and the similarity between the devices C-GaAs and E-GaAs demonstrates that the GaAs layer prevents the formation of extra QDs and that the properties of the devices are not only caused by the Al layer, but also by the core. From the present results, it cannot yet be concluded if the GaAs cover shell improves the intrinsic properties of a NW (higher mobility of carriers or ``harder'' superconducting gap). When Al etching is necessary, to use electrostatic barriers for example, the GaAs cover shell may be used to prevent the appearance of unwanted QDs without affecting the proximity effect. This will give the opportunity to focus in the future on the intrinsic properties of the wires using devices with electrostatic barriers or with NWs half-covered with Al. Although it is possible to form ohmic contacts directly on InAs, this extra protective shell should also be compatible with good contacts to the external leads made afterwards.
\\

\section{In-gap measurements}
We now present a study of a device similar to the device C-Al presented above, called C-Al'. The parameters of this device are listed in Table \ref{tab1}: the NW is smaller and the junctions are 1~$\mu$m from each other. The Al layer of both extremities of the NW (beyond the junctions) has been chemically etched. Note that in this paragraph, conductance (and not resistance) measurements are shown. The device displays regular Coulomb diamonds with a stability diagram similar to the one given in the insert of Fig. \ref{fig3}. Therefore we will focus in the subgap regime with $| eV_{bias}|\leq2(\Delta+E_c)$. Figure \ref{fig5} shows the conductance measurements at zero field with the theoretical model (see below). In panel \textbf{(a)}, the  experimental (on the left hand side) and modelled (right hand side) stability diagram show clear Coulomb features in the subgap regime and the panels \textbf{(b)} and \textbf{(c)} are the measurements at constant $V_{BG}$ and $V_{bias}$ respectively. The difference between the gate open to the gate close state is more visible in panel \textbf{(b)} and a clear dip close to $V_{bias}=0$ is present. Close to $V_{bias}=0$ the ratio of the conductances reaches $G_S/G_N\sim 10^{-6}$ and $G_S/G_N\sim 10^{-3}-10^{-4}$ at low bias outside the dip, highlighting the good quality of the proximized superconductivity. The panel \textbf{(c)} shows the Coulomb oscillations with the gate and their period doubling when $|V_{bias}|\leq50~\mu$V, i.e. when the bias voltage is smaller than the charging energy. These features - change of periodicity from 1$e$ to 2$e$ of the Coulomb oscillations and pronounced dip at low bias - are robust and have been observed in several devices. Similar pronounced minimum in the conductance as the one we observed has already been reported in proximized NWs \cite{Doh2008} and is attributed to the Coulomb blockade regime, but with a conductance ratio $G_S/G_N$ several orders of magnitude larger than in our device.

\begin{figure}
	\centering
		\includegraphics[width=0.75\textwidth]{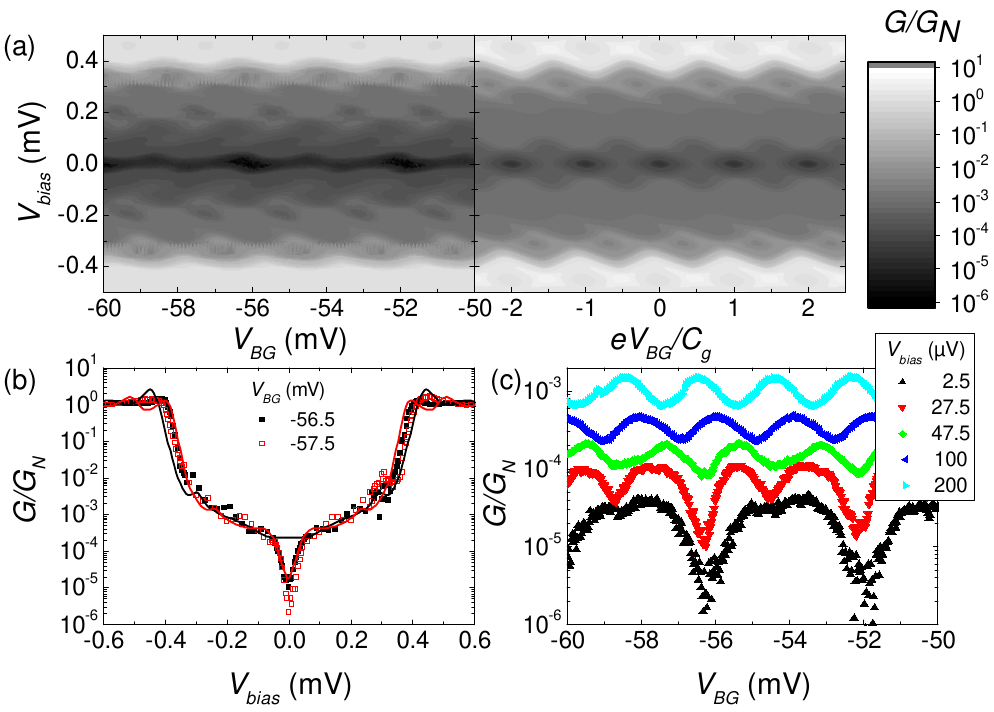}
	\caption{Conductance measurement of the device C-Al'. Panel \textbf{(a)}: Experimental (on the left) and theoretical (in the right) stability diagram with clear Coulomb features in the subgap regime. Panel \textbf{(b)}: Measurements at constant backgate voltage with the theoretical fits in solid lines. Panel \textbf{(c)}: Coulomb oscillation of the conductance with the bakgate at constant bias, a change of periodicity is occurring at $V_{bias}\approx50~\mu$V.
	}
	\label{fig5}
\end{figure}

The model of the conductance measurements shown Fig.~\ref{fig5} is similar to the simple one used in Fig.~\ref{fig3} taking into account the normalized Dynes density of states (DOS) in the superconducting state \cite{Dynes1978,Dynes1984}

\begin{equation}
	n_{\text{D}}(E)=\bigg| \Re e \frac{E/\Delta+i\gamma}{\sqrt{(E/\Delta+i\gamma)^2-1}}\bigg|
\end{equation}

with $\gamma$ the Dynes parameter. The model shown in Fig. \ref{fig5} uses $\gamma=6 \times 10^{-4}$ and the parameters in Table \ref{tab1}. It reproduces relatively well the measurements as shown Fig.~\ref{fig5}, except for the low bias voltage regime. Indeed, in this range the theoretical fit cannot reproduce the 2$e$ periodicity of the oscillations and tends to overestimate the conductance. The origin of the 2$e$-periodic signal and of the dip at low bias voltage is not clear yet, but they might come from localized in-gap states, which have been observed in similar devices using the same type of NWs \cite{Higginbotham2015}. A more advanced model is therefore be needed to describe fully our system, and devices with higher charging energy would be useful in order to disentangle accurately the effect of these potential in-gap states to the Coulomb blockade regime.

\begin{figure}
	\centering
		\includegraphics[width=0.75\textwidth]{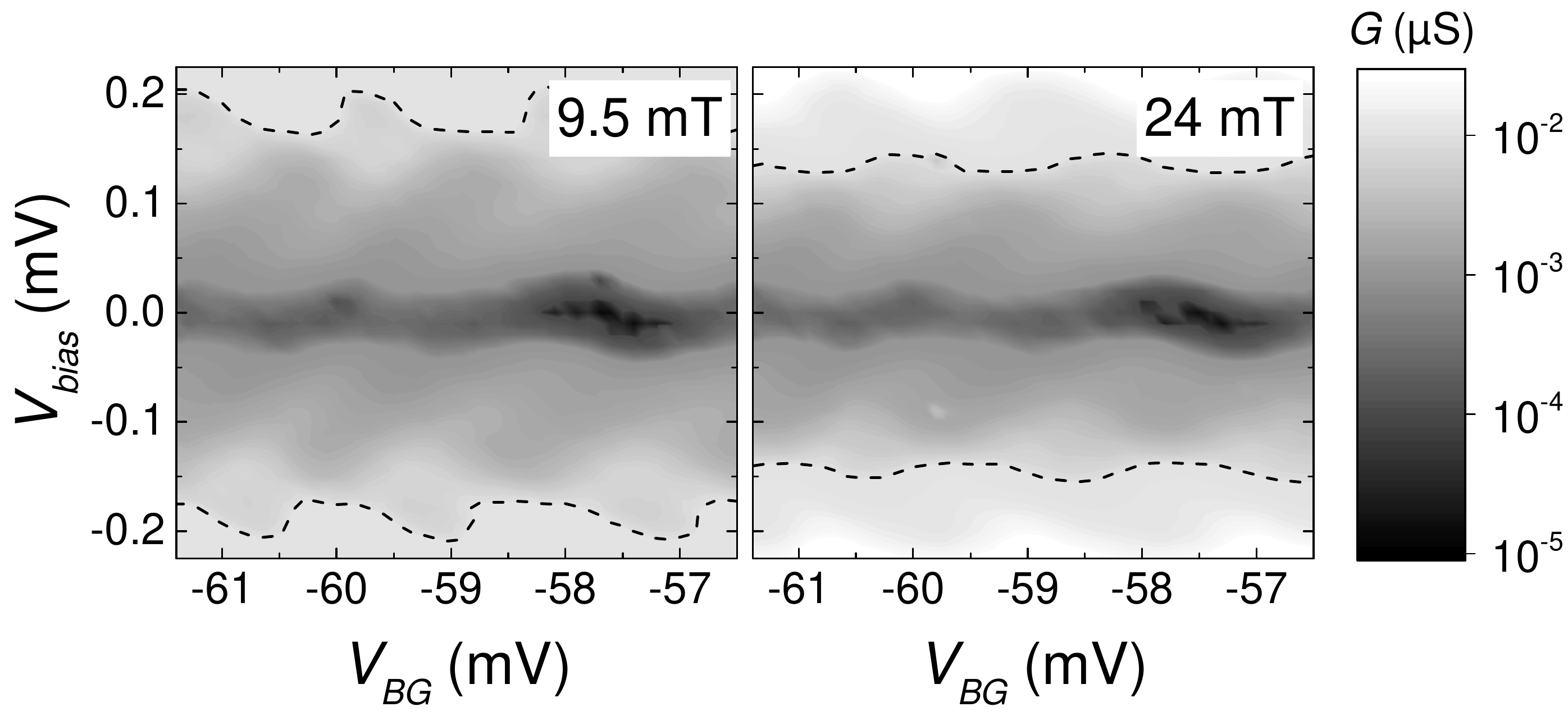}
	\caption{Stability diagram at low bias taken at 9.5~mT (on the left) and at 24~mT on the right hand side. The dashed lines are the iso-conductance lines at $G=10~$nS.
	}
	\label{fig6}
\end{figure}

Figure \ref{fig6} shows the stability diagram at low bias of the device C-Al' at $B=9.5~\text{mT}$ (left hand side) and $B=24~\text{mT}\simeq B_{c2}/2$ (right side). The main effect of the magnetic field is to reduce the superconducting gap, as shown by the iso-conductance lines (dashed lines in Fig. \ref{fig6} at $10~$nS) going closer to zero bias, the in-gap features at low bias voltage being relatively field insensitive below 30~mT. By increasing further the magnetic field, the 2$e$-periodic signal will eventually vanish and the superconducting gap will close completely. The complete study of the magnetic field and temperature dependence of the subgap features are needed to get a better understanding.

\section{Conclusion}
In conclusion we have demonstrated that InAs nanowires proximized with aluminium can be used as a single-electron transistor with a hard superconducting gap, by forming a fixed tunnel barrier based on the aluminium shell. Our results confirm that unwanted quantum dots can appear on the surface of the InAs core when bare. As for our devices the aluminium shell does not have to be etched, this prevents the formation of these extra quantum dots. They can be avoided when an additional thin protective layer of GaAs is inserted between the InAs and the aluminium, without seemingly degrading the transport or the superconducting properties of the system. Our technique provides a way to minimize the number of gates needed to nanowire-based devices. This gives an opportunity to use an InAs nanowire as an island of a single-electron transistor with the rich properties of a nanowire.

\subsection{Acknowledgements}
We thank M. Meschke and J. T. Peltonen for technical advice, and N. Paillet for help in plasma etching. 
This work has been supported by Academy of Finland (projects 272218 and 284594), by Danish National Research Foundation and by Microsoft Project Station Q.
 We acknowledge the availability of the facilities and technical support by Otaniemi research infrastructure for Micro and Nanotechnologies (OtaNano).

%


\end{document}